# Theoretical understanding of correlation between magnetic phase transition and the superconducting dome in high-$T_c$ cuprates

Chen Zhang[1,2], Cai-Xin Zhang[1], Su-Huai Wei[3,*], Haiqing Lin[3,*], Hui-Xiong Deng[1,2,*]

[1]*State Key Laboratory of Superlattices and Microstructures, Institute of Semiconductors, Chinese Academy of Sciences, Beijing 100083, China*

[2]*Center of Materials Science and Optoelectronics Engineering, University of Chinese Academy of Sciences, Beijing 100049, China.*

[3]*Beijing Computational Science Research Center, Beijing 100193, China*

Many issues concerning the origin of high-temperature superconductivity (HTS) are still under debate. For example, how the magnetic ordering varies with doping and its relationship with the superconducting temperature; and why the maximal $T_c$ always occurs near the quantum critical point. In this paper, taking hole-doped La$_2$CuO$_4$ as a classical example, we employ the first-principles band structure and total energy calculations and Monte Carlo simulations to explore how the symmetry-breaking magnetic ground state evolves with hole doping and the origin of a dome-shaped superconductivity region in the phase diagram. We demonstrate that the local antiferromagnetic ordering and doping play key roles in determining the electron-phonon coupling, thus $T_c$. Initially, the La$_2$CuO$_4$ possesses a checkerboard local antiferromagnetic ground state. As the hole doping increases, $T_c$ increases with the increase of the density of states at the Fermi surface. But as the doping increases further, the strength of the antiferromagnetic interaction weakens. At the critical doping level, a magnetic phase transition occurs that reduces the local antiferromagnetism-assisted electron-phonon coupling, thus diminishing the $T_c$. The superconductivity disappears in the heavily overdoped region when the antiferromagnetic ordering disappears. These observations could account for why cuprates have a dome-shaped superconductivity region in the phase diagram. Our study, thus, contributes to a fundamental understanding of the correlation between doping, local magnetic ordering, and superconductivity of HTS.

[*]**Corresponding authors. E-mail:** *suhuaiwei@csrc.ac.cn*, *haiqing0@csrc.ac.cn*, and *hxdeng@semi.ac.cn*

Even though high-temperature superconductivity (HTS) has been discovered more than thirty years ago, it still presents many challenging issues for the rich exotic phenomena [1-4] and continue to fascinate physicists. Compared to the conventional electron-phonon (*el-ph*) pairing theory, a key characteristic of HTS is being inherently and strongly correlated with the magnetic state, as has been confirmed by extensive theoretical and experimental studies in copper-, iron- and nickel- based superconductors [5-7]. The magnetic states play an essential role in determining high-temperature superconducting behaviors. For example, for the cuprate family, extensive experimental studies establish a consensus that the antiferromagnetic (AFM) ordering could coexist with superconductivity in the underdoped region [8-10]. As the doping level increases, the AFM interaction is suppressed, and a magnetic quantum critical point (QCP) occurs around the optimum doping level [11,12], that is confirmed by many experimental phenomena such as the divergent electronic specific heat capacity at zero temperature [13], the sudden jump of the carrier concentration [14], the change of Fermi surface topology [15], the linear resistivity [16], and recently has also been signaled by the disappearance of charge order [17]. On the other hand, theoretically the spin fluctuation pairing model [18,19] based on the AFM ordering also successfully addresses the d-wave pairing symmetry [20] and the strange metal behavior at optimum doping [21].

As doping further increases into the overdoped region, many experimental observations suggest the possibility of some new local magnetic configurations may exist. Through the muon spin resonance ($\mu$SR) measurements, MacDougall *et al.* have proposed that the clustered staggered magnetization ordering should exist [22], and such magnetic behavior has also been favored by the susceptibility measurements [23,24]. Li *et al.* reported that a $\sqrt{2}a_0 \times \sqrt{2}a_0$ charge order begins to dominate in the overdoped region [17], which could be the implication of a new magnetic ordering possibly with a period twice the charge ordering, similar to the 1/8 doping stripe phase situation [25]. Recently, Ma et al. have discovered that the static AFM order exists throughout the entire superconducting dome. They have found that the static magnetic order does not impede superconductivity, but rather acts as a prerequisite for it [26]. Furthermore, as doping further increases, the local ferromagnetic ordering is observed in the heavily overdoped non-superconducting samples [27-29], indicating the weak ferromagnetism deteriorates the superconducting behaviors beyond the dome region.

However, despite such a richness of magnetic states suggested by the experiments, little is known about the exact spin configuration of these exotic magnetic phases and how the magnetic ordering varies with doping and determines the superconducting temperature, as well as what is related to the quantum phase transition around the optimum doping. Therefore, it is highly desirable to clarify the evolution of magnetic orderings with doping, and how they determine the formation of the doping-dependent dome-shaped superconductivity region with distinct QCP in the phase diagram.

In this work, by taking hole-doped $La_2CuO_4$ (La-214) as a classical example for cuprates, we employ the first-principles band structure and total energy calculations and Monte Carlo simulations to explore how the magnetic ground state evolves with hole doping, and further utilize frozen-phonon calculations and band symmetry analysis to identify the superconducting behaviors in relation to the different magnetic ground states in cuprates. We find the magnetism-assisted *el-ph* coupling in local AFM backgrounds can cause a large band splitting around the Fermi surface due to symmetry reduction, and leads to a large increase in $T_c$. As the hole doping increases, the density of states (DOS) near the Fermi surface increases, so is the $T_c$, but the strength of the antiferromagnetic interactions decreases [30]. Around optimum doping near QCP, a magnetic phase transition is identified, resulting in the reduction of the local AFM assisted *el-ph* coupling strength, and $T_c$ decreases. The doping dependence of experimental $T_c$ in La-214 is, thus, naturally reproduced by the *el-ph* coupling under different magnetic ground states, and the dome-shaped superconductivity region with QCP in cuprates is demonstrated. Our study shows that the pairing mechanism in cuprates calls for a reassessment with careful consideration of strongly coupled lattice, charge and magnetism.

The undoped La-214 has a Cu-centered octahedron that is elongated in the *c*-axis direction with $D_{4h}$ symmetry, thus the degenerate $d_{xy}$ and $d_{x^2-y^2}$ orbitals split into two nondegenerate levels, and the highest lying $d_{x^2-y^2}$ orbital is half-filled for the Cu $d^9$ configuration [31], making each Cu atom has a magnetic moment of around 0.6 $\mu_B$, in well agreements with experiments [32,33]. Although it has been confirmed that the magnetic ordering of Cu atoms is in the checkerboard AFM configuration with dominated nearest-neighbor interaction in the undoped La-214 [34], the nearest-neighbor coupling strength diminishes with increasing doping concentration [30]. Therefore, to explore the evolvement of magnetic ground state with different hole doping, we searched the magnetic ground state by adopting a cluster expansion method coupled with first-principles total energy calculations and Monte Carlo simulations [35] (more details can be seen in Fig S1 in supplemental materials (SM)). Fig. 1a shows the fitted multi-spin interaction parameters. It is found that as the hole doping level increases, $J_{p1}$ (two-spin interaction between nearest-neighboring Cu sites) continuously decreases, which indicates the increasingly unfavorable antiparallel arrangement between the nearest-neighboring Cu sites, i.e., the nearest-neighboring AFM correlation is seriously compromised upon hole doping. This stems from the fact that when the spin-up $d_{x^2-y^2}$ state deviating from the full-filled occupation, the energy gain of the anti-parallel magnetic superexchange interaction decreases while the energy gain of the parallel direct magnetic interaction increases [30].

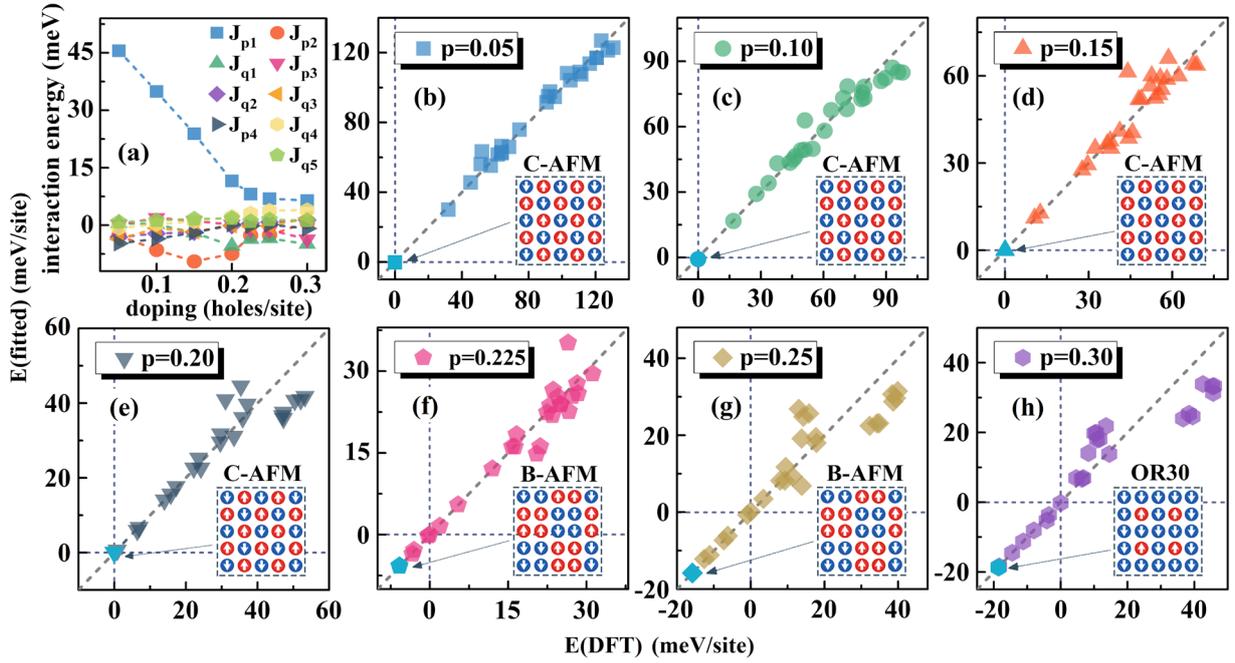

FIG. 1(a) Fitted multi-spin interaction parameters under different doping levels $p$. (b-h) Comparison between DFT calculated energy [E(DFT)] and predictions based on cluster expansion model [E(fitted)] for all magnetic configurations considered, E(fitted) is calculated using the multi-spin interaction parameters shown in (a). The Monte Carlo simulated ground states (GS) are marked by cyan symbols and its magnetic configurations are displayed in the insets. All energies are referenced to the energy of C-AFM configuration from DFT calculations. Note the change of scales in these plots.

The magnetic ground states for the different hole doping levels are obtained using the Monte Carlo simulation combined with the cluster expansion with parameters shown in Fig. 1a. Fig. 1b-h shows the magnetic configurations of the ground state for different doping levels, from the underdoped, optimum-doped, to the overdoped region in La-214. It is clearly shown that in the underdoped region, the magnetic configuration is the nearest-neighbor Checkerboard AFM (C-AFM, insets in Fig. 1b-e), which is consistent with experimental observations [34]. As the doping level increases, the nearest-neighbor C-AFM interaction gradually weakens (Fig.1a), but up to the optimum doping level (~0.20 holes/site), it still keeps the C-AFM ground state. Crossing the optimal doping point, a magnetic phase transition of the ground state occurs, i.e., it changes from C-AFM to Block AFM (B-AFM, insets in Fig. 1f). Further increasing the hole doping to the normal conducting region (~0.30 holes/site) is accompanied by another phase transition from B-AFM to OR30 (inset in Fig. 1h), which has a net magnetic moment in the Cu-O layer of La-214. The disappearance of C-AFM at around 0.20 holes/site coincides with a handful of experiments that indicate the disappearance of AFM glass state when above the optimum doping [8-10]. This, for the first time, provided theoretical evidence for the experimental speculation [11,12] that QCP could be related to magnetic phase transition.

In order to clarify the relationship between the magnetic phase transition and superconductivity in La-214, we examine the superconductivity for the different doping levels with different magnetic ground states using the frozen phonon method by taking full account of the on-site Coulomb repulsion $U$. The *el-ph* coupling constants are calculated based on the change of electronic eigenvalues around the Fermi surface, which can be approximately given by [36]: $\lambda = N(0) \sum_j \frac{1}{\omega_j} \langle [\Delta\varepsilon_{nk} - \Delta\varepsilon_F]^2 \rangle_{FS}$, where $j$ stands for different phonon modes, $N(0)$ is the total density of states (DOS) at the Fermi surface, $\Delta(u) = \langle [\Delta\varepsilon_{nk} - \Delta\varepsilon_F]^2 \rangle_{FS}$ is defined as averaged deformation potential at the Fermi surface, $\Delta\varepsilon_F$ represents the chemical potential change during lattice vibration and $\Delta\varepsilon_{nk}$ stands for the change of electronic eigenvalues of states with band index $n$ and momentum $k$, which is derived through shifting atoms by displacements of $u$. The displacement $u$ and the frequency $\omega$ are determined self-consistently to fulfill both $u = \sqrt{\frac{\hbar}{2M\omega}}$ and $E(u) = \frac{1}{2} M\omega^2 u^2$, where E($u$) is the corresponding potential energy curve. Because the *el-ph* coupling is dominated by the contributions of the phonons with a wave vector of ($\pi,\pi$) in cuprates [37], we choose these phonons for our frozen-phonon calculations in C-AFM configurations (Fig. 2a, Fig.2c). From the calculated deformation potentials of all considered phonon modes (more details presented in Fig S3 and Fig S5 in SM), the phonon mode that plays a dominant role in *el-ph* coupling is the oxygen full-breathing mode, which is also reflected in its large phonon anomaly (more details presented in Table S1 in SM). At a doping level of 0.05 holes/site, when non-magnetic (NM) state is assumed, phonon frequency of the full-breathing mode calculated from the potential energy curve is

67.5 meV, in good agreement with the value of 68.8 meV derived from the conventional DFPT calculation in Ref. [38], but deviates badly from the neutron-scattering data of 87.2 meV [39]. This large discrepancy vanishes when we choose the C-AFM configuration for Cu magnetic moments in the calculations. The calculated phonon frequency with the C-AFM configuration is 81.6 meV, which is much closer to the experimental value, implying the indispensable role of local AFM order in lattice dynamics.

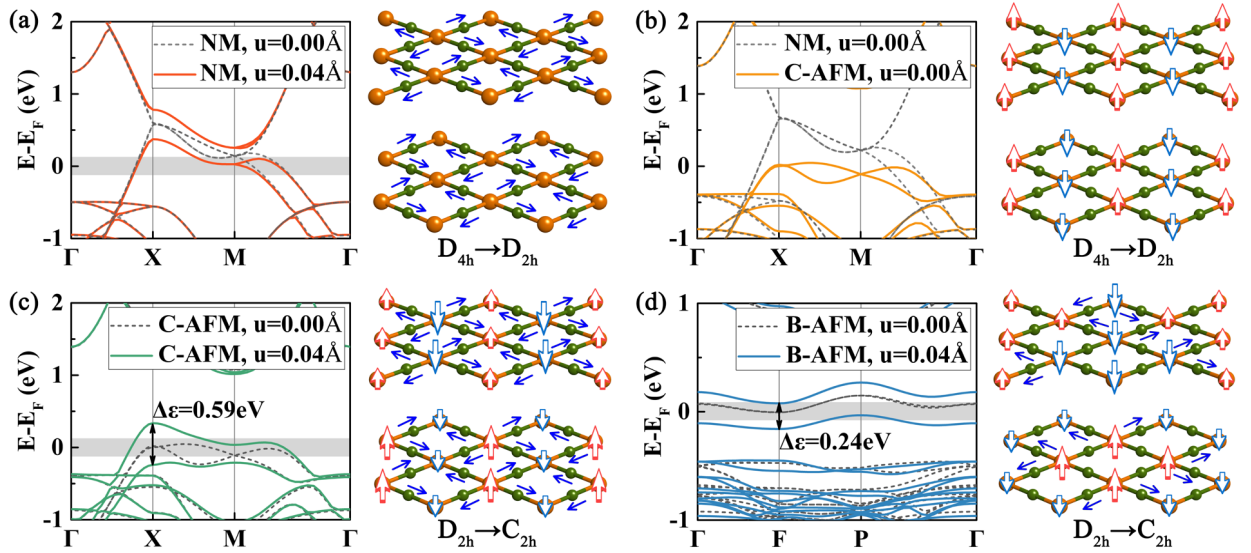

Fig.2. Comparison of band structures (a) before and after atomic displacements take place in NM phase; (b) C-AFM phase and NM phase with no atomic displacement; (c) before and after atomic displacements take place in C-AFM phase. (d) before and after atomic displacements take place in B-AFM phase. Since Kramers degeneracy still holds, only the spin-up part is presented. In all atomic configurations, Cu and O atoms are represented by the yellow and green balls respectively and all other atoms outside the Cu-O planes are omitted. The hollow arrows on the Cu atoms indicate the spin arrangement. Arrows alongside the O atoms denote the vibrations of phonon modes. All band structures in (a-c) are from La-214 with hole doping of 0.15 holes/site while 4 Cu atoms are included in supercells with 2 Cu atoms in each layer. The band structure in (d) is for La-214 with hole doping of 0.225 holes/site where 16 Cu atoms are included within a supercell with 8 Cu atoms in each layer.

Figure 2 shows the band structures of hole-doped La-214 with and without the oxygen-breathing displacements to clarify the effects of the *el-ph* coupling with the different magnetic configurations. When calculating band structures, the doping level is fixed at 0.15 holes/site and Hubbard $U$=6.0 eV is applied on Cu 3d orbitals. To include both C-AFM configuration and full-breathing mode vibration, we used a 28-atoms supercell with 4 Cu atoms distributed on the two identical Cu-O layers. In Fig. 2a, for the NM static structure, a four-fold degenerate point appears at X point due to band-folding twice because four equivalent Cu atoms are presented in the 28-atoms supercell (more details can be seen in Fig. S6 in the SM). Once oxygen breathing mode occurs, the Cu atoms in the same Cu-O plane are not equivalent, but the Cu atoms between the neighboring Cu-O planes are still equivalent, and the symmetry is reduced from $D_{4h}$ to $D_{2h}$. Therefore, the four-fold degeneracy at X splits into two two-fold degenerate points (Fig. 2a). However, because the isolated $d_{x^2-y^2}$ levels at the Fermi surface around X are still two-fold degenerate, and cannot produce a large crystal field splitting as atoms displace from their equilibrium positions for the NM phase, no significant energy level shift around the Fermi level is presented in Fig. 2a, resulting in poor *el-ph* coupling. We could conclude that phonons in NM phase play only a limited role in pairing. Such a view agrees with the previous conventional DFPT calculations [40-43], and is further confirmed by our calculated small deformation potentials (more details can be seen in Fig. S4 in SM) and minute *el-ph* coupling constants (Fig. 3a) when in NM phase. Consequently, the magnetic configurations should be taken into account in the *el-ph* coupling mechanism.

When considering the inequivalence between the nearest-neighbor Cu atoms with the anti-parallel magnetic moments, the C-AFM structure without atomic displacement also has a four-fold degeneracy to two-fold degeneracy splitting similar to the case with the displacement in the NM phase, resulting in a two-fold degenerate point around the Fermi level at X (Fig. 2b). When O atoms further vibrate in the C-AFM phase, the magnetic moments of Cu atoms with the increased plaquette areas (relatively large hollow arrows in Fig. 2c) are enhanced while the ones with the decreased plaquette areas are reduced (relatively small hollow arrows in Fig. 2c) in the breathing phonon mode. Hence, the adjacent Cu-O layers are no longer identical to each other because the anti-parallel net magnetic moments appear between neighboring Cu-O layers, thus the equivalency of the 4 Cu atoms within the supercell completely no longer exists and the symmetry is degraded to $C_{2h}$. The former two-fold degeneracy states further split into non-degenerate levels around the Fermi surface (Fig. 2c). Therefore, when C-AFM ordering is considered, the breathing phonon mode can cause a large band splitting around the Fermi surface (Fig. 2c). When the doping level reaches 0.15 holes/site, the band splitting at X is determined to be Δε=0.59 eV, signaling the much enhanced *el-ph* coupling when C-AFM correlation between Cu sites is included. Furthermore, when La-214 is overdoped to 0.225 holes/site where C-AFM magnetic ordering is transited into B-AFM, because the formerly discussed symmetry reduction still holds (when considering a phonon mode with eigenvectors match the magnetic order, as shown in Fig. 2d), the splitting around the Fermi surface still exists, but its value is reduced to Δε=0.24 eV at X, which indicates the *el-ph* coupling is compromised when in the overdoped region. Such a decrease of *el-ph* coupling strength when the system is overdoped has been experimentally observed previously [44] and agrees well with the phase diagram.

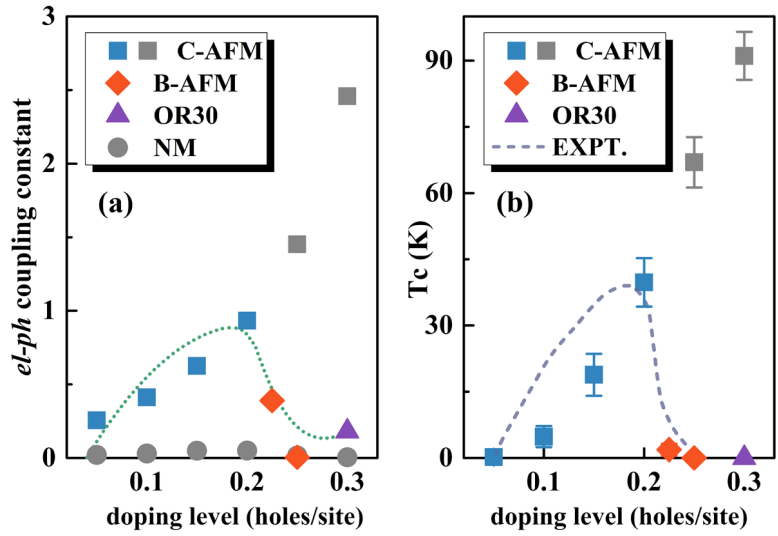

FIG. 3(a) The calculated *el-ph* coupling constants of the breathing mode in different magnetic backgrounds. (b) the calculated transition temperature ($T_c$) with all considered phonon modes included in different magnetic backgrounds. The upper and lower bounds of the error bars correspond to the effective Coulomb parameter taken as $\mu^*=0.10$ and $\mu^*=0.15$ respectively. Data calculated in a magnetic background that is not the ground state at the corresponding doping level are shown in grey symbols. The experimental data of single crystal La-214 is taken from Ref. [45].

Figure 3 presents the calculated *el-ph* coupling constants and the corresponding $T_c$ from Allen-Dynes modified McMillan equation [46]. It is clearly shown that the predicted $T_c$ with the different magnetic ordering has reached an overall agreement with the experiments. In the underdoped region where the C-AFM phase is sustained, the rise of $T_c$ is attributed to the increase in DOS at the Fermi surface when it gradually approaches the saddle point at M (more details can be seen in Fig. S7 in SM), which is confirmed by the calculated $N(0)$ increasing from 1.87 states/(eV·site) at 0.05 holes/site to 2.71 states/(eV·site) at 0.20 holes/site. At the same time, the enhancement of the doping also softens the breathing mode, causing a large displacement from the equilibrium positions, leading to a bigger perturbation potential, thus further boosting the *el-ph* coupling constants. However, when La-214 is doped into the overdoped region where B-AFM is formed, $N(0)$ and the deformation potential are both compromised, as $N(0)$ and the deformation potential are only 0.30 states/(eV·site) and $1.49 \times 10^{-4}$ eV$^2$ when the system is doped 0.25 holes/site, compared to 2.71 states/(eV·site) and 0.02 eV$^2$ when the system is doped 0.20 holes/site where C-AFM persists (more details can be seen in Fig. S7 in SM). As doping aggravates further, the in-plane time-reversal symmetry disappears due to the appearance of magnetic ordering with net magnetic moments, so the superconductivity is no longer expected.

In conclusion, using ab-initio calculations with Hubbard U, we have studied the doping dependence of magnetic states and the associated *el-ph* coupling in the dome-shaped superconductivity region in La$_2$CuO$_4$ cuprates. We demonstrate that the magnetism-assisted *el-ph* coupling in the original C-AFM phase can cause a large band splitting around the Fermi surface due to symmetry reduction, resulting in a high $T_c$. Initially, as the hole doping increases, $T_c$ increases with the increase of the density of states at the Fermi surface. As the hole doping increases further, the strength of nearest-neighboring antiferromagnetic interactions decreases. At the critical doping level, a magnetic phase transition occurs that weakens the local AFM assisted electron-phonon coupling, which leads to the reduction of $T_c$. The superconductivity disappears in the heavily overdoped region when the antiferromagnetic ordering disappears. This connection between doping, local magnetic ordering and superconductivity, thus, contributes significantly to the fundamental understanding of the physical mechanism of the AFM-related (e.g., copper- and iron-based) HTS.

This work was supported by the National Natural Science Foundation of China (Grants Nos. 61922077, 11874347, 11991060, 12088101, 61927901, U1930402), the National Key Research and Development Program of China (Grants No. 2018YFB2200100 and 2020YFB1506400), the Key Research Program of the Chinese Academy of Sciences (Grant No. XDPB22), and the CAS Project for Young Scientists in Basic Research (No. YSBR-026). H.-X. D. was also supported by the Youth Innovation Promotion Association of Chinese Academy of Sciences (Grant No. Y2021042).

# Supplementary Materials:

# Theoretical understanding of correlation between magnetic phase transition and the superconducting dome in high-$T_c$ cuprates

Chen Zhang[1,2], Cai-Xin Zhang[1], Su-Huai Wei[3]*, Haiqing Lin[3]*, Hui-Xiong Deng[1,2]*

[1]*State Key Laboratory of Superlattices and Microstructures, Institute of Semiconductors, Chinese Academy of Sciences, Beijing 100083, China*

[2]*Center of Materials Science and Optoelectronics Engineering, University of Chinese Academy of Sciences, Beijing 100049, China.*

[3]*Beijing Computational Science Research Center, Beijing 100193, China*

## Computational Methods

*First-principles calculations:* Our calculations employ density functional theory as implemented in Vienna Ab initio Simulation Package (VASP) [1,2] and the Perdew-Burke-Ernzerhof approximation [3] is utilized for the exchange-correlation functional. Following Ref. [4], although a tetragonal-to-orthorhombic transition is observed when doping level amounts p=0.20 holes/site at ambient pressure and 10 K, we described La-214 using tetragonal cells, which is a sensible simplification since studied of LSCO under pressure showed that at 20 Kbar the structure remains tetragonal for all doping levels, and at the same time the shape and magnitude of the superconducting dome is essentially identical to that measured at ambient pressure [5]. The cutoff energy of plane-wave basis is set to 400 eV. An on-site Coulomb repulsion *U* with value of 6.0 eV is applied on 3d orbitals of Cu atoms [6-8]. All original static structures were relaxed to their optimum configurations so that Hellman-Feynman forces on each atom reached less than 0.005 eV/Å. A Gaussian function with a smearing of 0.02 eV was employed for further statistics as a replica of the Dirac function.

*Cluster expansion method:* The total energies of different magnetic configurations are expanded by: $E(\sigma) = N \sum_f J_f D_f \bar{\pi}_f(\sigma)$, where $E(\sigma)$ is the total energy of the configuration $\sigma$ with *N* sites, $\bar{\pi}_f(\sigma)$ is the correlation function defined as the lattice averaged product of spins of figure *f* and $D_f$ is its degeneracy. For convenience, we assign different spin orientations as ±1. The multi-spin interaction parameters $J_f$ are determined by fitting the total energies of $n+1$ configurations (where $n$ is the number of considered clusters) with the smallest cell size at different doping levels. We gradually increase the number of clusters used for fitting based on the effective cluster interaction selection rule until the predicted ground state and the ground state from DFT calculations converge, meanwhile the cross-validation score is kept less than 0.025 eV/site [9]. Since the global $\mathbb{Z}_2$ symmetry holds for Ising spins in the absence of an external magnetic field, the total energy remains unchanged after all spins flip, thus the three-spin interaction terms should be zero and only the two-spin and four-spin interaction terms are kept, consistent with the *t/U* expansion of the Hubbard model [10,11]. Finally, 9 figures that are shown in Fig. S1 are determined to be the most sensible choice for the optimal cluster expansion construction.

*Monte Carlo simulations:* The ground states at different doping levels are obtained through Monte Carlo algorithm based simulated annealing. The 40×40 periodical square lattice is linearly cooled from 1000 K with a randomly generated fully disorder spin configuration to 0 K in a step of 2.5 K. At each temperature, the single-site Metropolis move is performed $40 \times 40 \times 100$ times to reach the corresponding thermally stable state. Furthermore, such an annealing process was performed 20 times at each doping level to prove the robustness of the derived ground states.

## OTHER NOTES

*Total energy competition between AFM and FM*: If the magnetic momentums between the nearest-neighboring Cu atoms have a parallel arrangement, the *d-d* coupling would not gain energy (Fig. S2(a)) since they are fully occupied in the spin-up channel, while the AFM configuration would result in an energy gain (Fig. S2(b)), thus La-214 is strongly AFM correlated when undoped makes sense. However, when each Cu site is doped with *p* holes, the energy gain when in FM configuration rises, while the energy gain when in AFM configuration decreases. Thus, the total energy in AFM configuration and FM configuration gets much closer compared to the undoped case, the local AFM correlation is much compromised when hole-doped.

*Band folding*: when a 14-atoms unit cell is built from 7-atoms primitive cell, the first Brillouin zone correspondingly shrinks, and the border perpendicular to the c-axis is replaced by the shaded area shown in Fig. S6(d), and all eigenstates in L-H-Q plane are folded onto Γ-M-K plane, thus points $X_1$

($\frac{\pi}{a}, \frac{\pi}{a}, \frac{2\pi}{c}$) and X$_3$ ($-\frac{\pi}{a}, -\frac{\pi}{a}, \frac{2\pi}{c}$) are folded onto points X ($\frac{\pi}{a}, \frac{\pi}{a}, 0$) and X$_2$ ($-\frac{\pi}{a}, -\frac{\pi}{a}, 0$) respectively (Fig. S6(b)). While due to the two-dimensional nature of $d_{x^2-y^2}$-$p$ hybridized orbitals in La-214, the band around the Fermi surface is poorly dispersive in c-axis, making points X and X$_2$ two-fold degenerate (Fig. S6(b)). Furthermore, when a 28-atoms supercell is built from 14-atoms unit cell, the in-plane first Brillouin zone shrinks from the purple square to the green square presented in Fig. S6(d), the original two-fold degenerate point X$_2$ is folded onto point X, hence point X becomes four-fold degenerate ultimately (Fig. S6(c)).

## SUPPORTING TABLES

Table S1 provides the comparison of calculated phonon frequencies of all considered phonon modes in NM and C-AFM configurations.

## SUPPORTING FIGURES

This includes figures from Fig. S1 to Fig. S7.

Fig. S1 presents all considered figures when cluster expansion method is performed.

Fig. S2 illustrates total energy competition between AFM and FM correlations upon hole doping can be explained by the *d-d* coupling.

Fig. S3 presents the eigenvectors of all considered phonon modes.

Fig. S4 presents the deformation potentials of all considered vibration modes when in NM configuration.

Fig. S5 presents the deformation potentials of all considered vibration modes when in C-AFM configuration.

Fig. S6 illustrates how the four-fold degenerate point at X in a 28-atoms supercell can be derived by band folding twice when in NM configuration.

Fig. S7 presents the calculated deformation potentials, the calculated DOS at the Fermi level and the evolution of the Fermi surface at $k_z$=0 upon hole doping.

Table S1: Comparison of calculated phonon frequencies at 0.05 holes/cell by potential energy curves with or without AFM background. the eigenvectors of each mode are displayed in Fig. S3.

| mode | ω (meV) | | | |
|---|---|---|---|---|
| | NM | C-AFM | DFPT (Ref. [12]) | Experiment (Ref. [13]) |
| $B_{3g}$ (rotational) | 19.5 | 16.6 | 8.5 | 17.0 |
| $B_{3u}$ | 18.4 | 19.0 | 17.7 | 19.0 |
| $B_{1g}$ | 28.1 | 27.0 | 22.3 | 26.9 |
| $B_{1u}$ | 41.5 | 41.7 | 38.9 | 40.0 |
| $A_{1g}$ | 56.0 | 56.6 | 52.4 | 60.4 |
| $B_{3g}$ (quadrupolar) | 63.1 | 63.0 | 54.5 | 62.4 |
| $A_{1g}$ (breathing) | 67.5 | 81.6 | 68.8 | 87.2 |

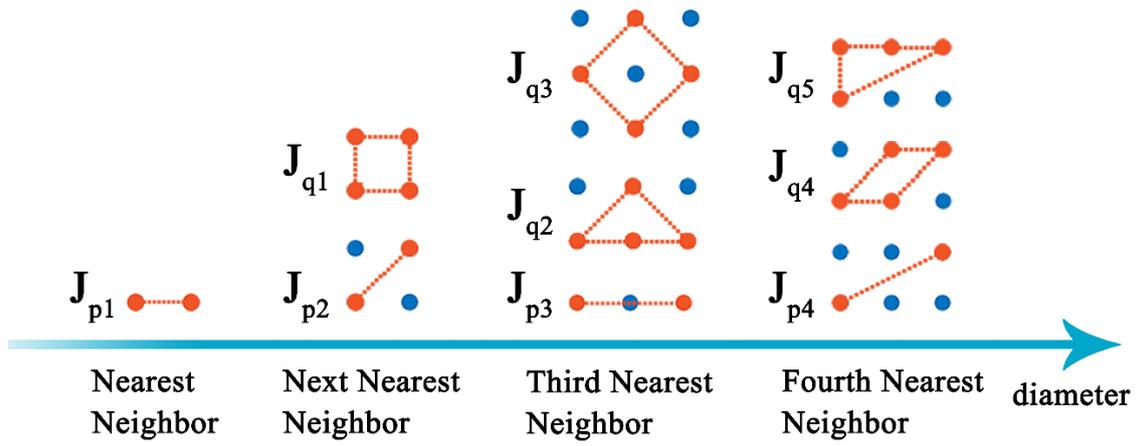

FIG. S1: Schematics of all considered figures when cluster expansion method is performed. A figure is defined by the number $k$ spins located on its vertices, which are denoted by the orange dots. In the subscripts, 'p' is for pairs and 'q' is for quadruplets.

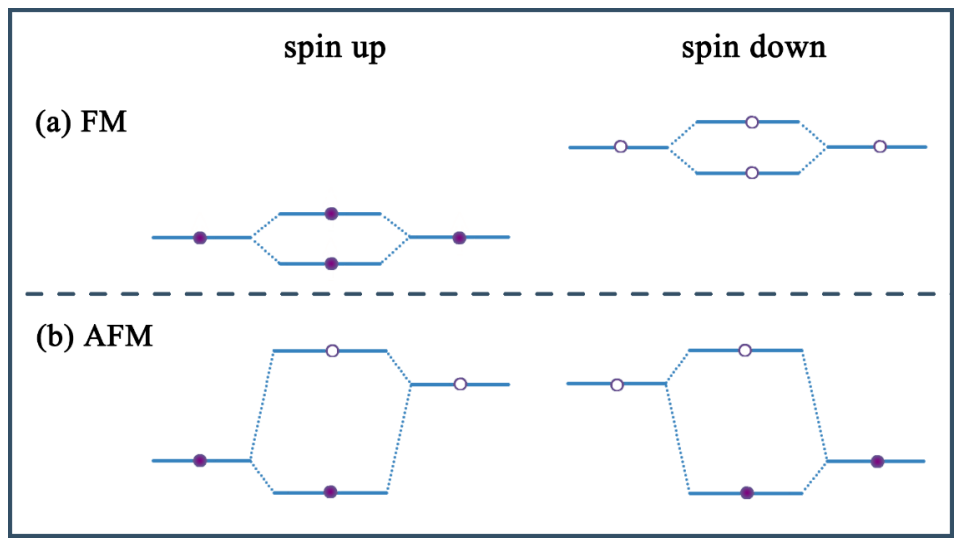

FIG. S2: Schematics of $d$-$d$ coupling in FM and AFM configurations.

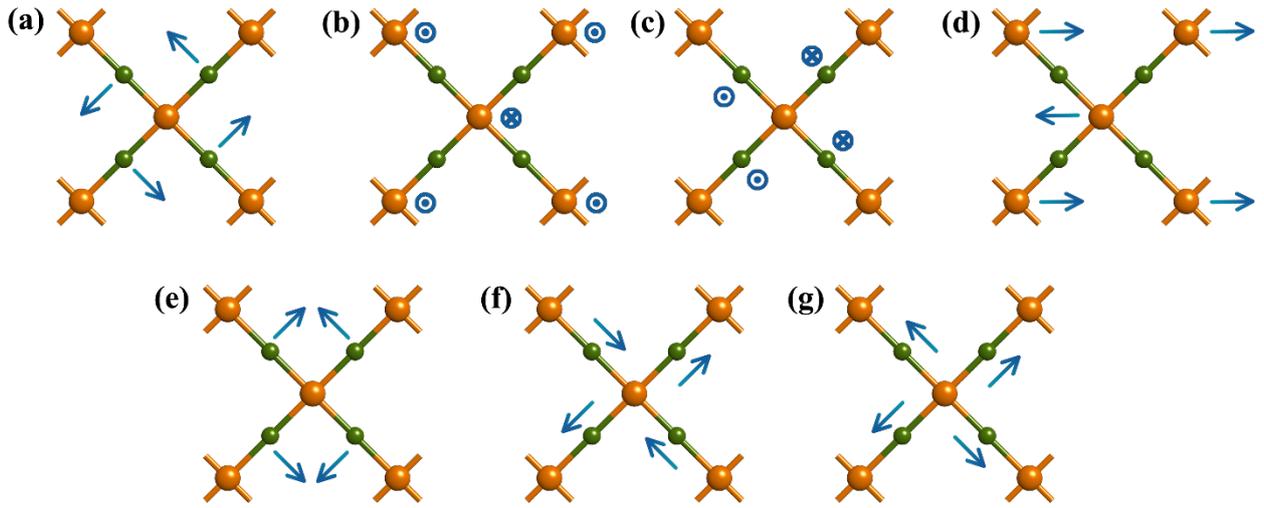

FIG S3: Schematics of all considered phonon modes at X point. (a) $B_{3g}$ rotational mode. (b) $B_{3u}$ Cu buckling mode. (c) $B_{1g}$ O buckling mode. (d) $B_{1u}$ mode. (e) $A_{1g}$ mode. (f) $B_{3g}$ quadrupolar mode. (g) $A_{1g}$ breathing mode. In all figures, yellow and green balls represent Cu atoms and O atoms respectively.

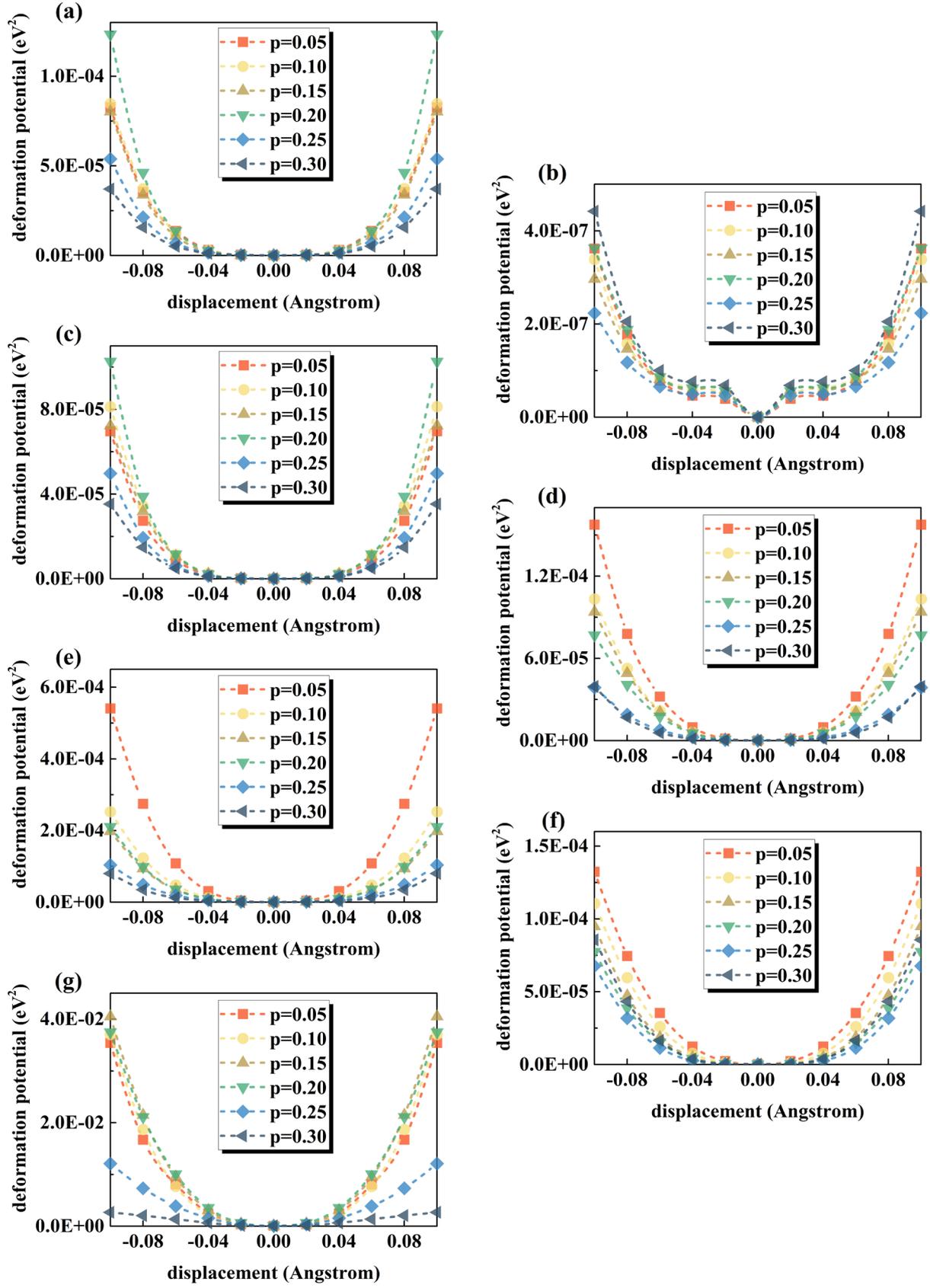

FIG S4: Calculated Fermi surface averaged deformation potential in NM background. The subplots in figure corresponds to the phonon modes presented in Fig. S3 in sequence.

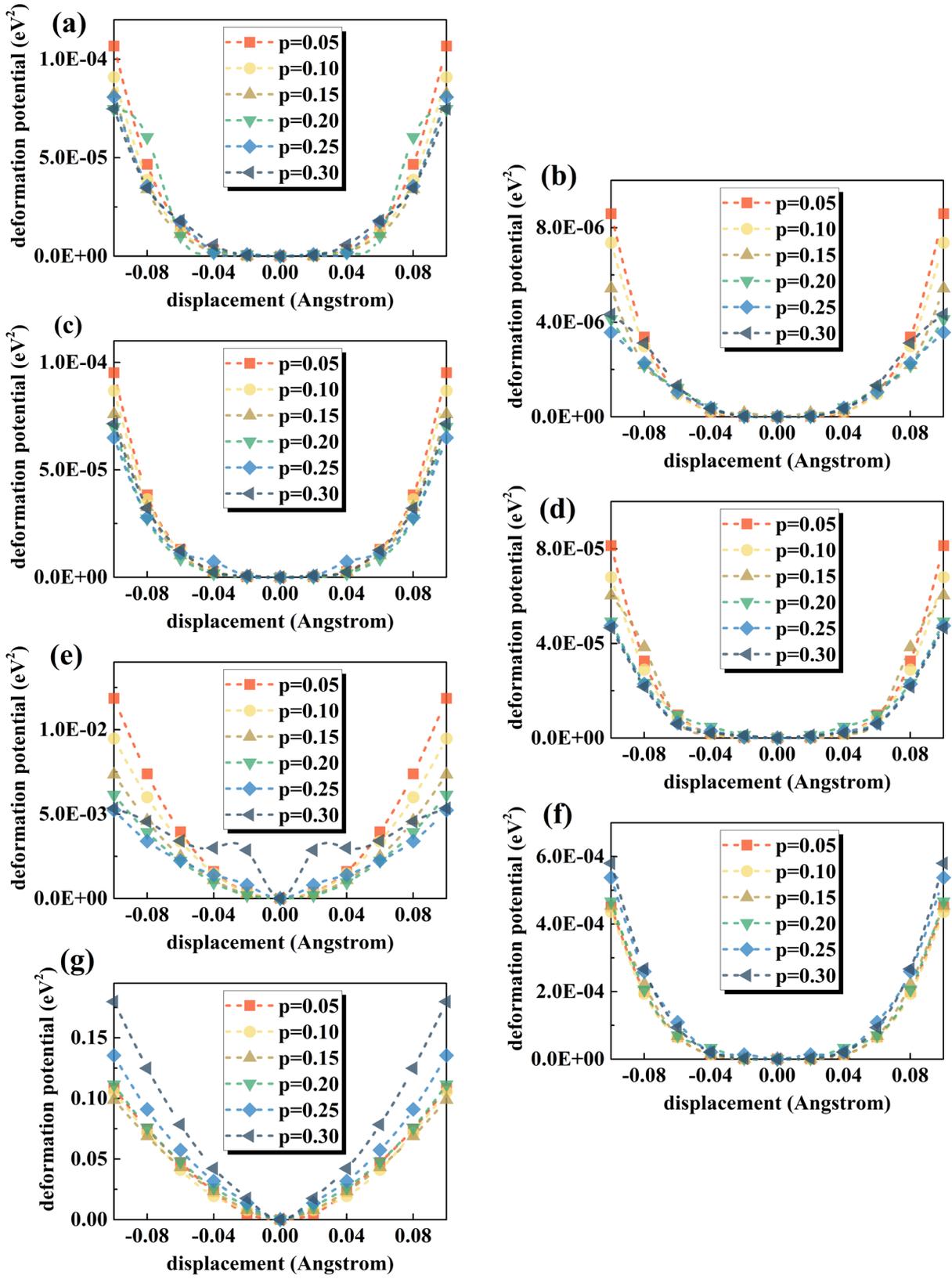

FIG S5: Calculated Fermi surface averaged deformation potential in C-AFM background. The subplots in figure corresponds to the phonon modes presented in Fig. S3 in sequence.

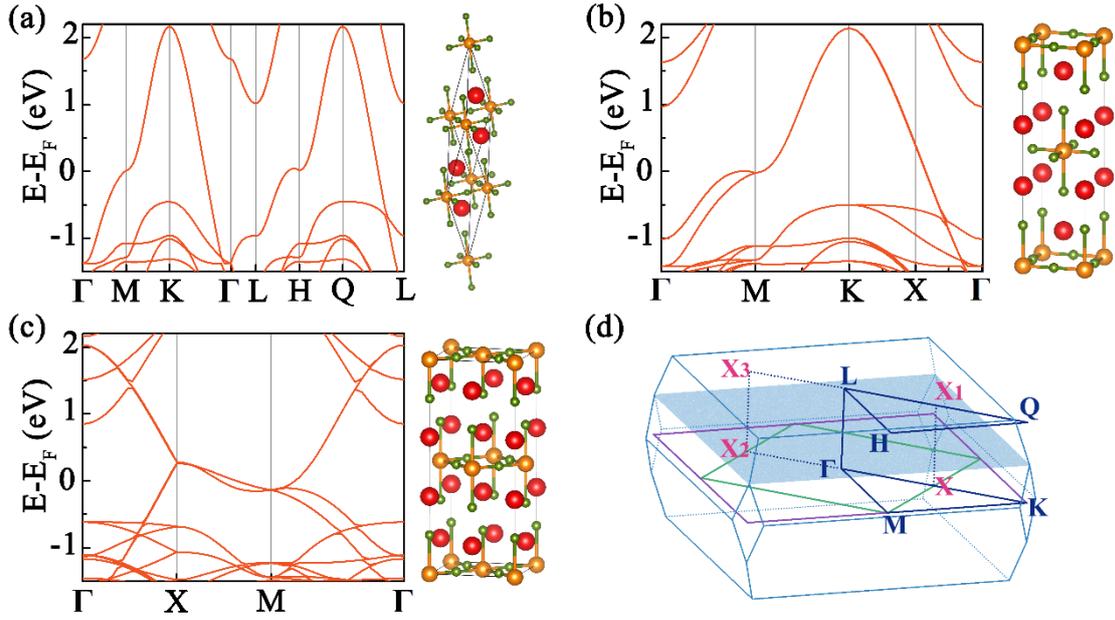

FIG S6: The process of band folding when the cell employed in calculations develops from (a) 7-atoms primitive cell to (b) 14-atoms unit cell, and further to (c) the ultimate 28-atoms supercell. The atomic structures and the corresponding band structures are shown in the corresponding subplots. The first Brillouin zone of the primitive cell is also shown in subplot (d). Red, yellow and green balls stand for La, Cu and O atoms respectively.

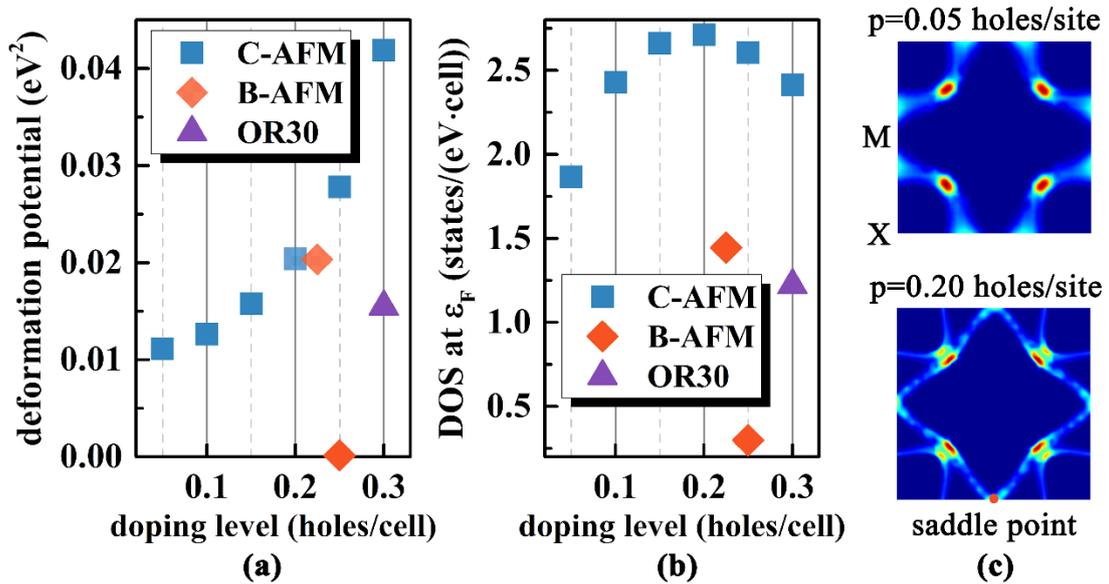

FIG S7: The calculated (a) deformation potentials (b) DOS at the Fermi surface and (c) the unfolded Fermi surface at $k_z=0$ in the C-AFM configuration, the saddle point at M is marked by the orange dot.